\documentclass[prl,preprint,superscriptaddress,showpacs,amssymb,amsmath,amsfonts,aps]{revtex4}
\usepackage{graphicx}
\usepackage{epsfig}
\usepackage{dcolumn}

\begin{document}

\title{Quark-Hadron Duality in Spin Structure Functions
$g_1^p$ and $g_1^d$}


\newcommand*{\ANL}{Argonne National Laboratory,  Argonne, Illinois 60439}
\affiliation{\ANL}
\newcommand*{\ASU}{Arizona State University, Tempe, Arizona 85287-1504}
\affiliation{\ASU}
\newcommand*{\UCLA}{University of California at Los Angeles, Los Angeles, California  90095-1547}
\affiliation{\UCLA}
\newcommand*{\CSU}{California State University, Dominguez Hills, Carson, CA 90747}
\affiliation{\CSU}
\newcommand*{\CMU}{Carnegie Mellon University, Pittsburgh, Pennsylvania 15213}
\affiliation{\CMU}
\newcommand*{\CUA}{Catholic University of America, Washington, D.C. 20064}
\affiliation{\CUA}
\newcommand*{\SACLAY}{CEA-Saclay, Service de Physique Nucl\'eaire, F91191 Gif-sur-Yvette, France}
\affiliation{\SACLAY}
\newcommand*{\CNU}{Christopher Newport University, Newport News, Virginia 23606}
\affiliation{\CNU}
\newcommand*{\UCONN}{University of Connecticut, Storrs, Connecticut 06269}
\affiliation{\UCONN}
\newcommand*{\ECOSSEE}{Edinburgh University, Edinburgh EH9 3JZ, United Kingdom}
\affiliation{\ECOSSEE}
\newcommand*{\FAI}{Fairfield University, Fairfield, CT 06824}
\affiliation{\FAI}
\newcommand*{\FIU}{Florida International University, Miami, Florida 33199}
\affiliation{\FIU}
\newcommand*{\FSU}{Florida State University, Tallahassee, Florida 32306}
\affiliation{\FSU}
\newcommand*{\GWU}{The George Washington University, Washington, DC 20052}
\affiliation{\GWU}
\newcommand*{\ECOSSEG}{University of Glasgow, Glasgow G12 8QQ, United Kingdom}
\affiliation{\ECOSSEG}
\newcommand*{\ISU}{Idaho State University, Pocatello, Idaho 83209}
\affiliation{\ISU}
\newcommand*{\INFNFR}{INFN, Laboratori Nazionali di Frascati, 00044 Frascati, Italy}
\affiliation{\INFNFR}
\newcommand*{\INFNGE}{INFN, Sezione di Genova, 16146 Genova, Italy}
\affiliation{\INFNGE}
\newcommand*{\ORSAY}{Institut de Physique Nucleaire ORSAY, Orsay, France}
\affiliation{\ORSAY}
\newcommand*{\BONN}{Institute f\"{u}r Strahlen und Kernphysik, Universit\"{a}t Bonn, Germany}
\affiliation{\BONN}
\newcommand*{\ITEP}{Institute of Theoretical and Experimental Physics, Moscow, 117259, Russia}
\affiliation{\ITEP}
\newcommand*{\JMU}{James Madison University, Harrisonburg, Virginia 22807}
\affiliation{\JMU}
\newcommand*{\KYUNGPOOK}{Kyungpook National University, Daegu 702-701, South Korea}
\affiliation{\KYUNGPOOK}
\newcommand*{\MIT}{Massachusetts Institute of Technology, Cambridge, Massachusetts  02139-4307}
\affiliation{\MIT}
\newcommand*{\UMASS}{University of Massachusetts, Amherst, Massachusetts  01003}
\affiliation{\UMASS}
\newcommand*{\MOSCOW}{Moscow State University, General Nuclear Physics Institute, 119899 Moscow, Russia}
\affiliation{\MOSCOW}
\newcommand*{\UNH}{University of New Hampshire, Durham, New Hampshire 03824-3568}
\affiliation{\UNH}
\newcommand*{\NSU}{Norfolk State University, Norfolk, Virginia 23504}
\affiliation{\NSU}
\newcommand*{\OHIOU}{Ohio University, Athens, Ohio  45701}
\affiliation{\OHIOU}
\newcommand*{\ODU}{Old Dominion University, Norfolk, Virginia 23529}
\affiliation{\ODU}
\newcommand*{\PITT}{University of Pittsburgh, Pittsburgh, Pennsylvania 15260}
\affiliation{\PITT}
\newcommand*{\RPI}{Rensselaer Polytechnic Institute, Troy, New York 12180-3590}
\affiliation{\RPI}
\newcommand*{\RICE}{Rice University, Houston, Texas 77005-1892}
\affiliation{\RICE}
\newcommand*{\Turkey}{Sakarya University, Sakarya, Turkey}
\affiliation{\Turkey}
\newcommand*{\URICH}{University of Richmond, Richmond, Virginia 23173}
\affiliation{\URICH}
\newcommand*{\SCAROLINA}{University of South Carolina, Columbia, South Carolina 29208}
\affiliation{\SCAROLINA}
\newcommand*{\JLAB}{Thomas Jefferson National Accelerator Facility, Newport News, Virginia 23606}
\affiliation{\JLAB}
\newcommand*{\UNIONC}{Union College, Schenectady, NY 12308}
\affiliation{\UNIONC}
\newcommand*{\VT}{Virginia Polytechnic Institute and State University, Blacksburg, Virginia   24061-0435}
\affiliation{\VT}
\newcommand*{\VIRGINIA}{University of Virginia, Charlottesville, Virginia 22901}
\affiliation{\VIRGINIA}
\newcommand*{\WM}{College of William and Mary, Williamsburg, Virginia 23187-8795}
\affiliation{\WM}
\newcommand*{\YEREVAN}{Yerevan Physics Institute, 375036 Yerevan, Armenia}
\affiliation{\YEREVAN}
\newcommand*{\NOWJLAB}{Thomas Jefferson National Accelerator Facility, 
Newport News, Virginia 23606}
\newcommand*{\NOWOHIOU}{Ohio University, Athens, Ohio  45701}
\newcommand*{\NOWUNH}{University of New Hampshire, Durham, New Hampshire 03824-3568}
\newcommand*{\NOWUMASS}{University of Massachusetts, Amherst, Massachusetts  01003}
\newcommand*{\NOWMOSCOW}{Moscow State University, General Nuclear Physics Institute, 119899 Moscow, Russia}
\newcommand*{\NOWMIT}{Massachusetts Institute of Technology, Cambridge, Massachusetts  02139-4307}
\newcommand*{\NOWURICH}{University of Richmond, Richmond, Virginia 23173}
\newcommand*{\NOWODU}{Old Dominion University, Norfolk, Virginia 23529}
\newcommand*{\NOWCUA}{Catholic University of America, Washington, D.C. 20064}
\newcommand*{\NOWGEISSEN}{Physikalisches Institut der Universit\"{a}t Giessen, 35392 Giessen, Germany}
\newcommand*{\NOWLANL }{ Los Alamos National Laboratory, Los Alamos, New Mexico 87545}


\author{P.E.~Bosted}
     \email{bosted@jlab.org}
     \thanks{Corresponding author.}
\affiliation{\JLAB}
\author {K.V.~Dharmawardane} 
\altaffiliation[Current address:]{\NOWJLAB}
\affiliation{\ODU}
\author {G.E.~Dodge} 
\affiliation{\ODU}
\author {T.A.~Forest} 
\affiliation{\ODU}
\author {S.E.~Kuhn} 
\affiliation{\ODU}
\author {Y.~Prok} 
\altaffiliation[Current address:]{\NOWMIT}
\affiliation{\VIRGINIA}
\author {G.~Adams} 
\affiliation{\RPI}
\author {M.~Amarian}
\affiliation{\ODU}
\author {P.~Ambrozewicz} 
\affiliation{\FIU}
\author {M.~Anghinolfi} 
\affiliation{\INFNGE}
% ...
\collaboration{The CLAS Collaboration}
     \noaffiliation


%
%
\date{\today}


\begin{abstract}
New measurements of the spin structure functions of the proton
and deuteron $g_1^p(x,Q^2)$ and
$g_1^d(x,Q^2)$ in the nucleon resonance region are compared with
extrapolations of target-mass-corrected 
next-to-leading-order (NLO) QCD fits to higher energy data. 
Averaged over the entire resonance region ($W<2$ GeV),
the data and QCD fits are in good agreement in both magnitude
and $Q^2$ dependence for $Q^2>1.7$ GeV$^2/c^2$.
This ``global'' duality appears to result from
cancellations among the prominent ``local'' resonance regions: in
particular strong $\sigma_{3/2}$ contributions in the  
$\Delta(1232)$  region appear to be compensated by strong
$\sigma_{1/2}$ contributions in the resonance 
region centered on 1.5 GeV.
 These results are encouraging for the extension
of NLO QCD fits to lower $W$ and $Q^2$ than have been used previously.
\end{abstract}
\maketitle

The theoretical description of particle 
interactions has utilized quark-gluon degrees of freedom
at high energies and hadronic degrees of freedom at
low energies. With suitable averaging over resonant excitations,
the two approaches have been found in several cases 
to be nearly equivalent, a phenomenon referred to as 
quark-hadron duality. These cases include
$e^+e^-$ annihilation, semi-leptonic decays of heavy mesons,
electron-pion scattering, semi-inclusive deep-inelastic
scattering, and both spin-averaged and spin-dependent
inclusive lepton-nucleus scattering~\cite{PhysRep}, the subject of the
present investigation. Pragmatically, understanding
the limitations and applicability of quark-hadron
duality in this process is useful in order to define 
 the kinematic region in
which parton distribution functions (PDF) can be reliably extracted.

In lepton-nucleon scattering, the low and high energy regimes have
traditionally been separated using $W$, the invariant mass of the 
hadronic final state, and $Q^2$, the four 
momentum transfer squared.
A region of prominent nucleon resonances
is observed for $W<2$ GeV and $Q^2<10$ GeV$^2$/$c^2$, 
while for higher $W$ or $Q^2$ there is no longer
any obvious resonance structure.
Historically, quark-hadron duality was first observed 
in 1970 by Bloom and Gilman  ~\cite{Bloom1970} 
in the spin-averaged lepton-nucleon process. They noted that the
inclusive structure function, $F_2(W,Q^2)$ averages smoothly 
at low $W$ and $Q^2$ to the scaling function $F_2(W,Q^2)$
measured at high energy, using an empirical scaling variable
in place of the original Bjorken $x$ scaling variable.
Subsequently, Georgi and Politzer~\cite{Georgi1976} found
that quark-hadron duality is exhibited down to 
$Q^2 \sim$ 1 GeV$^2$/c$^2$
using the Nachtmann~\cite{Nachtmann1975}
scaling variable 
$\xi \equiv 2x/(1+\sqrt{1+4M^2x^2/Q^2})$
which approximates the purely kinematic higher twist
corrections arising from the non-zero nucleon mass $M$.
More recently, explicit target-mass (TM) corrections have
been derived~in the framework of QCD for both unpolarized~and polarized
structure functions~\cite{Blumlein98} that obviate the need for
an approximate scaling variable. 

To explain quark-hadron duality theoretically, de R\'ujula, Georgi,
and  Politzer~\cite{Rujula1975}
employed a perturbative operator product expansion of QCD
structure function moments. In this framework, quark-hadron
duality implies a small net effect from
higher twist contributions, once the kinematic target-mass 
contribution is taken into account. In a simple QCD picture, the
additional higher twist contributions (which are proportional
to powers of $1/\sqrt{Q^2}$) are due to quark-quark and
quark-gluon correlations. 
Close and Isgur~\cite{CloseIsgur2001} provided an interesting
explanation in the constituent quark model in terms of
cancellations from resonance contributions with opposite parity.
A recent theoretical QCD study~\cite{Liuti2003} of both
polarized and unpolarized structure functions incorporates
many of these concepts, with the addition of a 
careful treatment of so-called high-$x$ resummation corrections.
The authors  concluded that 
higher-twist corrections are suppressed more for 
the unpolarized structure function $F_2$ than for the 
proton polarized structure function $g_1^p$, 
where a sizable negative contribution is observed.
A comprehensive review of quark-hadron duality from both the
experimental and theoretical perspective was also 
published recently~\cite{PhysRep}.

Unpolarized structure function data exhibit
excitation-energy-averaged scaling not 
only averaged over the entire resonance
region ($M<W<2$ GeV), referred to as ``global duality'', 
but also in each of several restricted regions
in $W$, corresponding to the three prominent resonance regions
centered on $W=1.23$, 1.5 and 1.7 GeV, 
a phenomenon
referred to as ``local duality''. This was demonstrated experimentally
using high-statistical-accuracy data from Jefferson 
Lab~\cite{Niculescu1999}, and interpreted theoretically by
Carlson and Mukhopadhyay~\cite{Carlson1998} using the 
expected pQCD $Q^2$-dependence of nucleon  
transition form factors. 

The new data presented in 
this report augment previously available results for $g_1^p$
from SLAC~\cite{Baum,E143}, DESY~\cite{HERMES2002} 
and JLab~\cite{Fatemi2003,Yun2003} with higher statistical precision
and an expanded range of $Q^2$. This allows us to experimentally
examine local duality for $g_1^p$ much more accurately than
was previously possible. The addition of a considerable
body of deuteron $g_1^d$ data allows the first examination
of the isospin dependence of global duality in $g_1$. 

When testing duality there is an
intrinsic uncertainty as to which DIS curves to use for
comparison with the 
averaged resonance region data. In this paper, we choose
the average of two representative Next-to-Leading Order (NLO) 
QCD fits~\cite{AACfit,GSRVfit}
to polarized structure function data above the resonance region. The 
NLO evolution is considered to be reasonably reliable down to
$Q^2$ values of order 1 GeV$^2/c^2$. We choose to use fits
with NLO evolution, rather than LO or purely empirical fits to data,
to give the best possible estimate of the $Q^2$-dependence of
$g_1$. The high-energy data used in the NLO QCD
fits have relatively large errors compared to those
for unpolarized structure functions, particularly at high 
values of $x$ that tend to correspond to our resonance region 
data. Since precise error bands in our 
kinematic region are not available,
we ascribe a very approximate relative error 
of 10\% (20\%) to the $g_1^p$ ($g_1^d$) DIS fits, independent of $x$.
The error on 
the average deuteron DIS fit
is larger due to the much larger relative contribution of
negatively polarized 
quarks in the neutron compared to the proton.
We take kinematic target-mass corrections into account
using the prescription of Bl\"umlein and Tkabladze~\cite{Blumlein98}:
\begin{eqnarray}
g_1^{TM}(x,Q^2)&=&\frac{x}{\xi(1+\gamma)^{3/2}}g_1^{QCD}(\xi,Q^2) \label{equ:TMcor} \\
&+&  \frac{(x+\xi)\gamma}{\xi(1+\gamma)^2}
\int_{\xi}^{1} \frac{du}{u} g_1^{QCD}(u,Q^2)\nonumber \\
&-& \frac{\gamma (2-\gamma)}{2(1+\gamma)^{5/2}}
\int_{\xi}^{1} \frac{du}{u} \int_{u}^{1}\frac{dv}{v} g_1^{QCD}(v,Q^2),\nonumber 
\end{eqnarray}
where $\gamma = 4M^2x^2/Q^2$. This prescription is not unique,
and in particular has the drawback of resulting in non-zero
values of $g_1$ at $x=1$. An approach that avoids this problem
has been worked out for $F_2$~\cite{Wally2006}, but is not
yet available for $g_1$. The calculation of high-$x$ resummation
corrections is theoretically more complicated~\cite{Forte} and technically
more challenging than target mass corrections. Rather than 
attempting these calculations ourselves, we simply note
that Ref.~\cite{Liuti2003} finds enhancements of order 10\% to
20\% for the proton averaged over the full resonance region, 
roughly independent of $Q^2$ for $0.5<Q^2<5$ GeV$^2/c^2$. 
This corrections could well be different for the deuteron and
for individual ``local'' regions in $W$.

The analysis is based on recently published data~\cite{Vipuli2006} 
from Jefferson Lab. Very briefly, in this experiment 
the CEBAF Large Acceptance 
Spectrometer~\cite{CLASapp} in Jefferson Lab's Hall B 
was used to measure spin asymmetries in the scattering of 
longitudinally polarized electrons from 
longitudinally polarized protons and deuterons.
The data were collected in  2001 using incident  energies
of 1.6 GeV and 5.7 GeV. Beam currents ranged from 1 to 5 nA,
and the beam polarization averaged 70\%.
The detector package~\cite{CLASapp}
allowed clean identification of electrons scattered at polar
angles between 8 and 45 degrees. Ammonia, polarized via 
Dynamic Nuclear Polarization~\cite{Keith}, was used
to provide polarized protons and deuterons, using 
the $^{15}$NH$_3$ and  $^{15}$ND$_3$ isotopes, respectively.
The average target polarization was about 75\% for the
proton and about 25\% for the deuteron. The data were divided
into 40 bins in $Q^2$, equally spaced on a logarithmic scale
between 0.01 and 10 GeV$^2/c^2$.

\begin{figure}[ht]
\vspace{0.2in}
\scalebox{1.0}[1.0]{\includegraphics{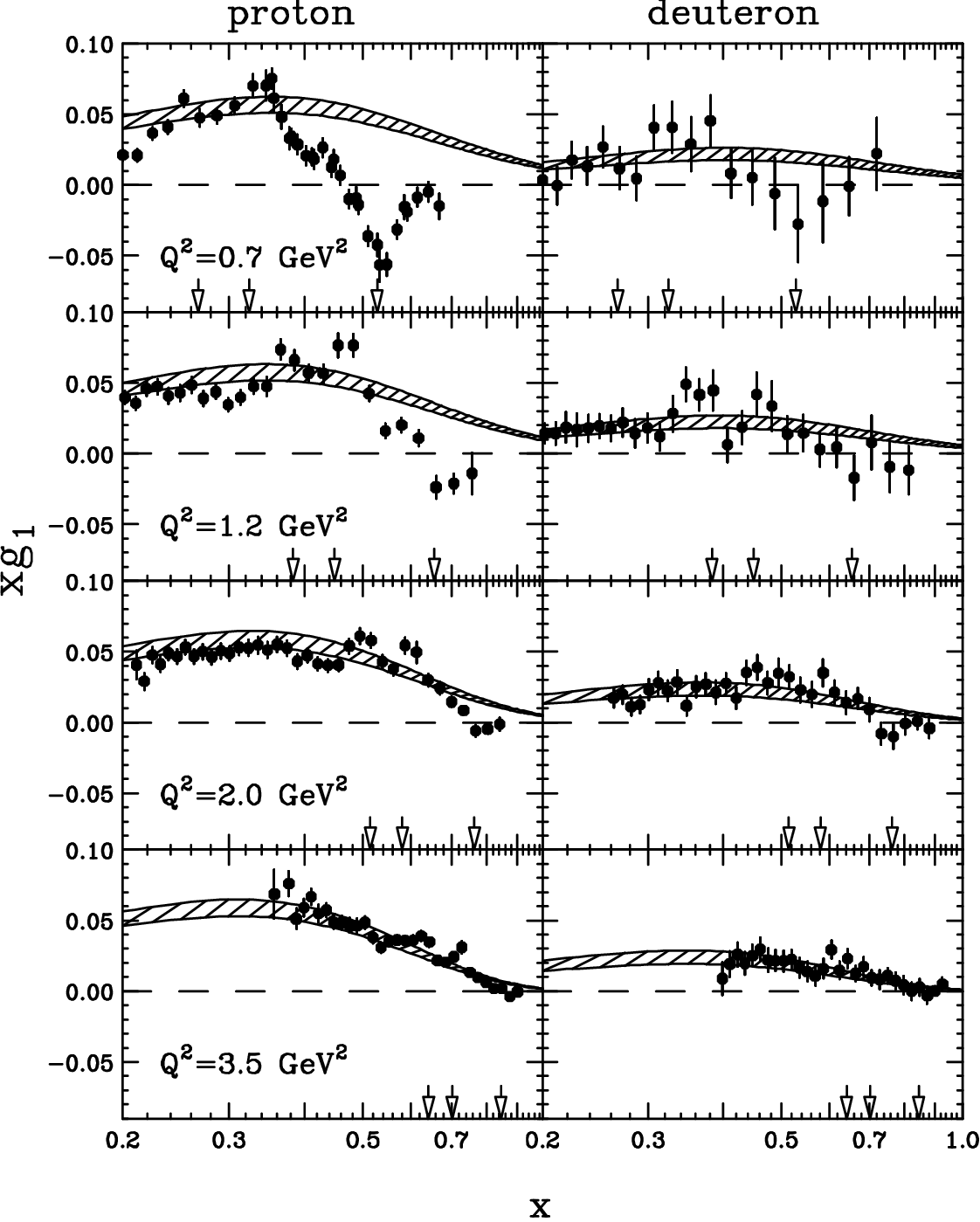}}
\caption{Present data for proton $g_1^p(x,Q^2)$ (left panels) 
and deuteron $ g_1^d(x,Q^2)$ (defined to be per nucleon, 
right panels) at four representative values of $Q^2$. 
The errors include statistical and systematic 
contributions added in quadrature. 
The three arrows on each plot indicate the central kinematic
position of the 
three prominent resonance regions at $W=1.7$, 1.5, and 1.23 GeV
from left to right. 
The hatched band
represents the range of $g_1$ predicted by modern NLO 
Parton Distribution Function (PDF) fits 
(GRSV~\protect{\cite{GSRVfit}} and
AAC~\protect{\cite{AACfit}}) to high energy data, 
evolved to the $Q^2$ of
our data and corrected for target-mass as
described in the text.}
 \label{fig:g1-vs-xi}
\end{figure}

Values of $g_1(x,Q^2)$ were determined from the ratios of 
$g_1/F_1$ presented in~\cite{Vipuli2006}
using recent fits to proton~\cite{F1pfit} and deuteron~\cite{F1dfit}
data to evaluate the unpolarized structure function $F_1(x,Q^2)$.
The resulting values of $g_1(x,Q^2)$ 
for both the proton and the deuteron are plotted 
(scaled by $x$) as a function of $x$ for four
representative $Q^2$ bins in Figure~\ref{fig:g1-vs-xi}.
The three arrows on each panel correspond to 
the three prominent resonance regions at 
$W=1.7$, 1.5, and 1.23 GeV, from left to right. We compare the
data to the extrapolations of DIS fits (as described above), 
represented by the hatched bands. 

It can be seen in Fig.~1 that the data in fixed $Q^2$ bins
indeed exhibit oscillations in $x$ compared to the smooth behavior
of the DIS curves. In addition, the averaged data become increasingly 
commensurate with the models with increasing $Q^2$, 
as expected if quark-hadron duality is valid for $g_1$. 
In closer detail, one can also observe that
the experimental data for both the proton and the deuteron
lie below the curves in the 
$\Delta(1232)$ region, and above in the $W=1.5$ GeV 
[$S_{11}(1535)$/$D_{13}(1520)$/$P_{11}(1440)]$ region.
This is not surprising at low to moderate $Q^2$, where resonant 
contributions dominate over non-resonant contributions.
Recall that $g_1$ is proportional
to $\sigma_{1/2}-\sigma_{3/2}$. The $N\rightarrow \Delta(1232)$ 
is known to be dominated by M1 strength~\cite{M1ref} over the $Q^2$
range of the present study, which 
results in  a virtual photon cross section $\sigma_{3/2}$ 
about three times larger than $\sigma_{1/2}$, corresponding to 
negative values of $g_1$.  In contrast, in the 
DIS limit of incoherent scattering
from massless quarks, $g_1$ must be positive due to helicity
conservation. In the second resonance region, the 
$S_{11}(1535)$ and $P_{11}(1440)$ transitions can
only contribute to $\sigma_{1/2}$, and recent studies~\cite{Burkert}
show that the $N\rightarrow D_{13}(1520)$ changes from dominantly
$\sigma_{3/2}$ at low $Q^2$ to dominantly $\sigma_{1/2}$ 
above 1 GeV$^2/c^2$. The three resonances together therefore are
expected to result in large 
positive values of $g_1$ above 1 GeV$^2/c^2$
(potentially larger than the DIS limit).

\begin{figure}[ht]
\vspace{0.4in}
\scalebox{1.0}[1.0]{\includegraphics{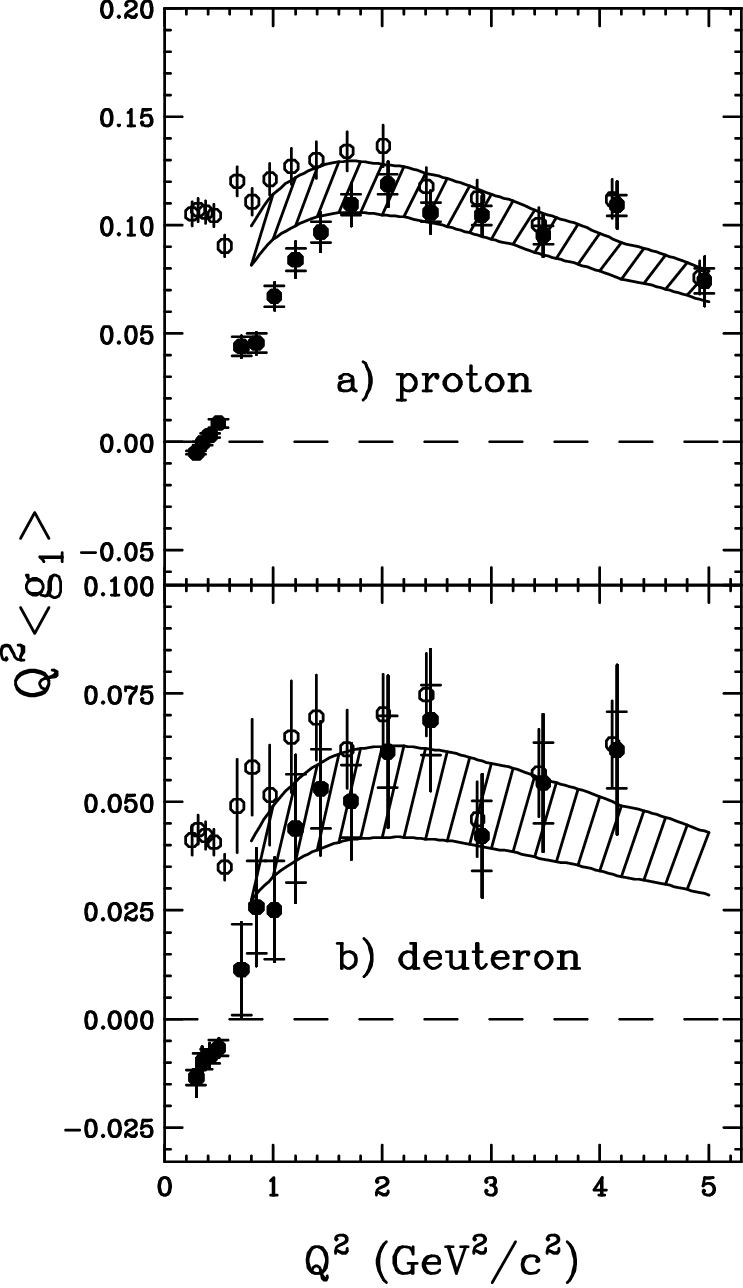}}
\caption{
The $Q^2$-dependence of $Q^2g_1(x,Q^2)$, averaged over 
a region in $x$ corresponding to $1.08<W<2$ GeV (solid circles)
for: a) proton; b) deuteron.
The inner error bars reflect only statistical
contributions, while the outer error bars  include 
statistical and systematic components added in quadrature. 
The open circles represent our
data after adding the contribution from $ep$ elastic ($ed$ quasielastic)
scattering at $x=1$ for the proton (deuteron). 
For clarity these results are slightly displaced in $Q^2$,
and the error bars include only statistical contributions. 
The hatched bands represent 
the range of the averages calculated from extrapolated NLO
DIS fits, as in Fig.~1 (see text for details).}
\label{fig:RatioTestp}
\end{figure}

\begin{figure}[ht]
\vspace{0.4in}
\scalebox{1.0}[1.0]{\includegraphics{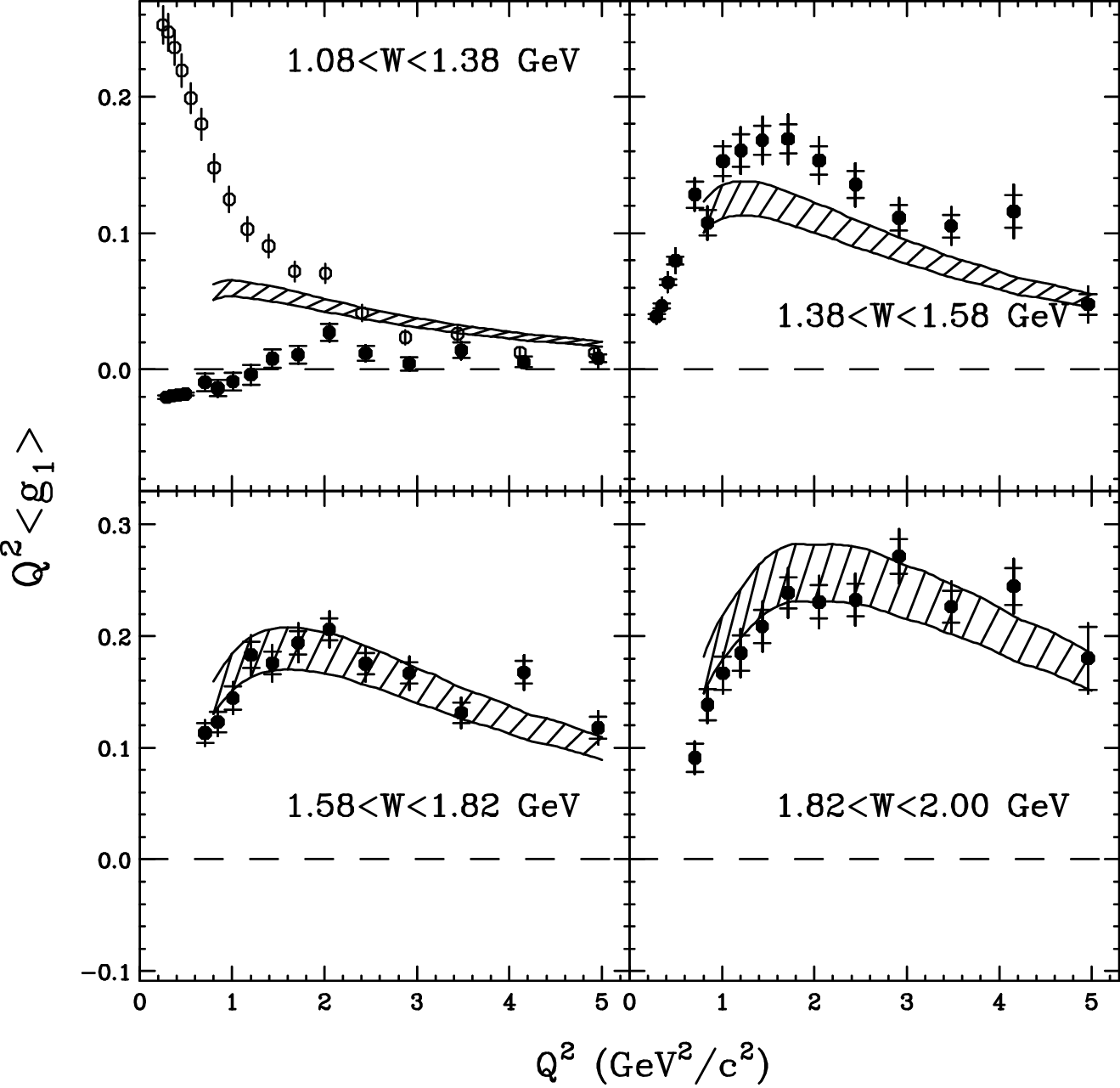}}
\caption{
The $Q^2$-dependence of $Q^2g_1(x,Q^2)$ for the proton,  averaged over 
various regions of $x$.
At each $Q^2$, the $x$-range over which $g_1$ was
averaged is determined by the corresponding range in $W$ as indicated
in each panel (see text). Symbols and curves as in 
Fig.~\protect{\ref{fig:RatioTestp}}.}
\label{fig:RatioTestd}
\end{figure}

To clarify these observations with respect to both local and global
duality, we have averaged over $x$ both data and models 
for $g_1$ over a $Q^2$-dependent interval corresponding to 
four specific regions in $W$. 
The $x$-averaged values of $g_1$ for the entire resonance region
(scaled by $Q^2$) are plotted
as a function of $Q^2$  in 
Fig.~\ref{fig:RatioTestp} for both targets.
The proton averages for four smaller regions in $W$ are plotted in
Fig.~\ref{fig:RatioTestd}.
Specifically, the averages
were determined as
$$<g_1(Q^2)>={\int_{x_l}^{x_h}g_1(x,Q^2)dx \over x_h-x_l},$$ \noindent where
$x_l$ and $x_h$ correspond respectively to the maximum and minimum values
of $W$ in the interval considered,
at the given value of $Q^2$ [using the definition
$x^{-1}=1+(W^2-M^2)/Q^2$]. 
The TM-corrected NLO PDF parameterizations
shown in Fig.~1 were averaged in the same way as for the 
experimental data. 

The averages displayed in 
Figs.~2 test ``global'' duality by averaging $g_1$ over $x$ for
the entire region from pion threshold to $W=2$ GeV. The data for
both targets exhibit a power-law-type deviation 
from the DIS curves at low $Q^2$, but essentially agree
with them above $Q^2=1.7$ GeV$^2/c^2$, within the systematic
errors of the data and models. This is somewhat higher in
$Q^2$ than for the unpolarized $F_2$ structure, supporting the 
conclusions of Ref.~\cite{Liuti2003}.


Turning to the examination of ``local'' duality for the proton,
it can be seen in the upper left panel of Fig.~3 that in the
``first'' resonance region, dominated by the $\Delta(1232)$
resonance, the data have the opposite sign of the extrapolations
of DIS models at low $Q^2$. This is allowed by
the spin-3/2 nature of the $\Delta(1232)$, and is expected due to the
dominance of the M1 transition strength~\cite{M1ref}, as discussed above.
What is interesting is that, while the data change sign at
$Q^2$ near 1 GeV$^2/c^2$, the averaged values 
are significantly below the models even to the highest
$Q^2$  of the present experiment, in spite of the fact that 
the $N\rightarrow \Delta(1232)$ transition form factor (FF) decreases
more rapidly with $Q^2$ than, for example, the elastic FF or the 
$N\rightarrow S_{11}(1535)$ transition FF~\cite{Stoler,M1ref}
 (a phenomena sometimes referred to as the 
``disappearing $\Delta$''). It is evident that the $\Delta(1232)$
has not yet completely disappeared at $Q^2=5$ GeV$^2/c^2$. 
 
In the second resonance region, two of the three known resonances
[$P_{11}(1440)$ and $S_{11}(1535)$] contribute only to $\sigma_{1/2}$,
while the third [$D_{13}(1520)$] contributes more to $\sigma_{1/2}$ than to 
$\sigma_{3/2}$ above 1 GeV$^2/c^2$~\cite{Burkert}. Therefore,
it isn't surprising that the proton data lie significantly above the
DIS extrapolations in this narrow region of $W$.

In the ``third'' resonance region centered on 1.7 GeV, 
the $F_{15}(1680)$ is dominant at low $Q^2$, but above about
1 GeV$^2/c^2$ other resonances are also important~\cite{Burkert}, such as
the $S_{11}(1650)$, $S_{31}(1620)$, and $D_{33}(1700)$.  
The $F_{15}$ contributes mainly to $\sigma_{3/2}$ at low $Q^2$ 
(i.e. negative $g_1$), but switches to $\sigma_{1/2}$ 
dominance at higher $Q^2$~\cite{Burkert}.
The average over all of these resonances plus non-resonant background
produces very good agreement between data and DIS models
in this region, as might be expected from the parity-averaging
arguments of Close and Isgur~\cite{CloseIsgur2001}.

For completeness, we have also studied a fourth region centered
on 1.9 GeV (for which there are a large number of poorly
established resonances, difficult to distinguish from non-resonant
contributions). In this case, the data lie slightly below the
DIS models, although the significance is marginal when systematic
errors are taken into account. 
It appears that much of the good agreement between data and models
observed in the entire resonance region comes about from
pairing the ``first'' and ``second'' resonance regions together,
with further improvement from including the ``fourth'' region. 
This lends further support to the Close-Isgur model.

Again following Close and Isgur~\cite{CloseIsgur2001}, one might
expect DIS and resonance region data to converge at lower
values of $Q^2$ if the ground state elastic contribution
is also included in the global duality averaging. 
The open circles in Fig.~2
include the elastic (quasi-elastic) contributions to the
$g_1$ averages  for the proton (deuteron), given by 
$G_E(Q^2)[G_E(Q^2) + \tau G_M(Q^2)]/2(1+\tau)(x_h-x_l)$, 
where $\tau=Q^2/4M^2$.
To evaluate the nucleon electric and
magnetic form factors
$G_E(Q^2)$ and $G_M(Q^2)$, we used a slightly modified version of the 
parameterization of Ref.~\cite{ne11fit}.
For both the proton and the deuteron, the $Q^2$-dependence 
with the elastic contribution more closely resembles the
$Q^2$-dependence of the models, down 
to values of $Q^2$ as low as 0.7 GeV$^2/c^2$, which is already
pushing below the expected region of validity of 1 GeV$^2/c^2$
for the PDF fits. However, the magnitude of the data is of 
order 10\% to 20\% higher than the models. As mentioned above,
high-$x$ resummation corrections may well account for this
difference. 
The result of pairing the $\Delta(1232)$ resonance
(the lowest spin-3/2 ground state) with the elastic
contribution (the lowest spin-1/2 ground state) is illustrated in
the upper left panel of Fig.~3 for the proton. The
elastic contribution, in the way we have treated it, over-compensates,
resulting in power-law deviations at low $Q^2$ that lie well
above the data, rather than well below. It appears that including
the elastic contribution with the entire resonance region works
much better than pairing it with the single $\Delta(1232)$ resonance.

In summary, we have used data for both the proton and deuteron
to examine both ``local'' and ``global''  
quark-hadron duality in $g_1$. As was determined in previous
studies~\cite{Liuti2003,PhysRep}, $g_1^p$ in the resonance region 
oscillates around extrapolations of NLO PDF fits to higher
energy data, especially when target-mass corrections are taken
into account. Averaged over the traditional resonance region
($W<2$ GeV), the data and fits agree within errors for $Q^2>1.7$
GeV$^2/c^2$, a slightly higher value than observed for the spin-averaged
structure function $F_2$~\cite{Liuti2003,Niculescu1999}.
Including the elastic contribution may extend the region of 
agreement to $Q^2=0.7$ GeV$^2/c^2$, after consideration of the
uncertainties due to high-$x$ resummation.
A similar effect was found in the unpolarized 
case~\cite{PhysRep,Niculescu1999}. We find similar results for
the previously unexamined $g_1^d$ structure function as for $g_1^p$,
indicating no large effects from different isospin projections. 
In terms of
``local'' duality, we find that in the $\Delta(1232)$ region, 
the proton data lie below the PDF fits, even for $Q^2$ values as large
as 5 GeV$^2/c^2$, while 
the region centered on 1.5 GeV lies above the
PDF fits for all $Q^2$ values studied. It appears that 
global duality is largely realized by summing over the four lowest-mass
resonances.  
Since
the $Q^2$-dependence for the PDF fits and the data are 
remarkably similar above $Q^2=1.7$ GeV$^2/c^2$, we conclude that
from the practical point of view, it is not unreasonable, with
suitable averaging over $W$, to
include data with $W>1.58$ GeV and $Q^2>1.7$ GeV$^2/c^2$ in future
global NLO PDF fits, as long as the effects of TM and high-$x$
resummation effects are taken into account. In the near future,
high statistical accuracy data from the present experiment
with 2.5 and 4.2 GeV electron energies, and additional data at 5.7 GeV,
presently under analysis,  will allow more precise
studies, particularly for the deuteron. 

We would like to acknowledge the outstanding efforts of the 
Accelerator, Target Group, and Physics Division staff that made
this experiment possible. 
This work was supported by the U.S. Department of Energy, the
Italian Istituto Nazionale di Fisica Nucleare, the U.S. National
Science Foundation, the French Commissariat \'{a} l'Energie Atomique 
and the Korean Engineering and Science 
Foundation. The Southeastern
Universities Research Association (SURA) operates the Thomas
Jefferson National Accelerator Facility for the United States
Department of Energy under contract DE-AC05-84ER40150.

\end{document}